\input epsf.tex
\documentclass[12pt]{article}
\begin{document}

\begin {titlepage}

\begin{flushright} ULB-TH/02-08\\   hep-th/0203096\\
\end{flushright}

\begin{center} {\large \bf  A Brief Course in Spontaneous Symmetry 
Breaking \\ I.  The Paleolitic Age}\footnote{Invited
talks presented at the 2001 Corfu Summer Institute on Elementary 
Particle Physics}\\
\vspace{1cm}  Robert Brout
\\
\vspace{.5cm} {\it Service de Physique Th\'eorique}\\ {\it
Universit\'e Libre de Bruxelles, Campus Plaine, C.P.225}\\ {\it
Boulevard du Triomphe, B-1050 Bruxelles, Belgium}\\
\end{center}

\begin{abstract} 

\noindent
The physical word is marked by the phenomenom of spontaneous broken
symmetry (SBS) i.e. where the state of a system is assymmetric with
respect to the symmetry principles that govern its dynamics. For
material systems this is not surprising since more often than not
energetic considerations dictate that the ground state or low
lying excited states of many body system become ordered i.e. a
collective variable, such as magnetization or the Fourier transform
of the density of a solid, picks up expectation values which
otherwise would vanish by virtue of the dynamical symmetry(isotropy
or translational symmetry in the aforementioned examples). More
surprising was the discovery of the role of SBS in describing the
vacuum or low lyng excitations of a quantum field theory.  First came
spontaneously broken chiral symmetry which was then applied to soft
pion physics. When combined with current algebra,  this field dominated
particle physics in the 60's. Then came the application of the notion of
SBS to situations where the symmetry is locally implemented by gauge
fields. In that case the concept of order becomes more subtle. This
development lead the way to electroweak unification and it remains
one of the principal tools of the theorist in the quest for physics
beyond the standard model. This brief review is intended to span
the history of SBS with emphasis on conceptual rather than
quantitative content. It is a written version of lectures of
R.Brout on the ``Paleolithic Age''  and
on ``Modern Times'' by F.Englert,  i.e. respectively without and with gauge
fields.

\end{abstract}
\end {titlepage}
\addtocounter{footnote}{-1}
\parskip 10pt plus 1pt
\setlength{\parindent}{0cm} 

{\bf \large I. The Early Ancestors (van der Waals and Weiss)}\cite{brout}
\medskip

At the beginning of the 20th century, van der Waals proposed the
idea of a ``molecular field" in order to explain the deviation of the
equation of state of gases from ideality and from there to
condensation. (What this has to do with SBS will emerge
subsequently.) His idea was to consider that each molecule was
surrounded by others which interact with it. Thus its energy is
\begin{equation}V_{mol}=\left[ \int d^{3}r^{\prime }v(r-r^{\prime
})\right] \rho\, , 
\end{equation}
where $v( r)$ is the intermolecular potential, taken to be
attractive, in van der Waals's eyes, for $r>r_{0}$; $\rho$ is the mean
density. The ``molecular field approximation'' (MFA) is to neglect
the correlation of density at $r^\prime$ to the presence of a molecule at
$r$. Though this neglect does some injustice to the situation, we have
learnt over the years that, in the large, the essential physics is
respected. One exception is the quantitative theory of critical
phenomena,     so beautifully executed by Wilson, Fisher and
others. However, throughout this review we shall work in MFA since
the main progress which has been made in analyzing the order
encountered in a great variety situations has been in MFA. (Once
more a notable exception is in 2 dimensional systems wherein
topological considerations are often vital). In general as the
dimensionality increases so does the reliability of MFA and for
$d>4$ , it becomes reliable in all thermodynamic conditions (In
this review we shall not touch upon lattice gauge theory where
dimensionality plays a different role from the more conventional
many body and field theoretic systems treated here. Thus confinement
will not be included).

From Eq.(1), van der Waals deduced the existence of an internal
pressure, $p_{int}$ given by
\begin{equation} 
p_{int}=-\frac{\partial V_{mol}}{\partial v}=\rho
^{2}
\frac{\partial V_{mol}}{\partial \rho }=\rho
^{2}\widetilde{v}(o)\, ,
\end{equation}
where $\widetilde{v}(o)$ is the Fourier transform of $v(r)$ at $q=0$. The
total pressure is thus $p+p_{int}$; $p$ is the external pressure. Under
normal conditions ($ p\cong 1atm$ and $T\cong $ room temperature) a
typical liquid exhibits $ p_{int}=10^{3}atm$, which gives one an
idea of just how essential are the intermolecular forces in
maintaining the cohesion of the liquid, as against a vapor where
$p_{int}$ more often than not is negligible away from critical
conditions.

Whereas in an ideal gas has $p=kT/v$ ($v$ being the volume per
molecule), van der Waals proposed that in a general fluid
one should replace $v$ by the ``free volume'', that which is unoccupied
by the molecule itself. Thus he set $p_{total}=(kT/v-b)$\ where $b$
is volume occupied by stuff within a single molecule. He thus set
\begin{equation}
p_{Tot}=p+\rho ^{2}\widetilde{v}(o)=kT[\rho
^{-1}-b]^{-1}\, ,
\end{equation}
the famous van der Waals equation. This equation of state has been
qualitatively successful but fails quantitatively near the critical
point, as is to be expected. In Fig.1 we sketch schematically a few 
isotherms
\vskip .5cm
\hskip 2.5cm
\epsfbox{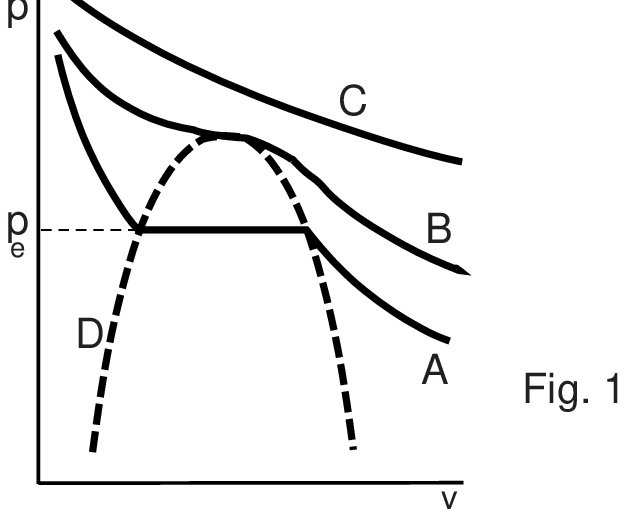}

\begin{quote}
\small
Only vapor exists for all $p$ once $T\geq T(c)$, but
for
$T<T(c)$, along, for example, isotherm A, the system is in the
liquid (vapor) phase for $p>p(e)$ ( $ p<p(e)$~)\ respectively. At
$p=p(e)$, the system has a choice between liquid and vapor. They
can coexist. It
is this choice, seemingly arbitrary, which in   this case is SBS.
We shortly bring more clarity into this question, but for the
moment we ask the reader to bear in mind the coexistence curve D,
which on the left side marks the locus of where liquid isotherms
begin as the pressure increases and on the right where vapor
isotherms take over as the pressure decreases.
\end{quote}

\bigskip

\normalsize 

Whilst van der Waals was busy in Holland explaining why it rains or
shines, Weiss in France was proposing a similar mechanism to explain
ferromagnetism. Like van der Waals, he was toying with this new
fangled idea of the atomic hypothesis. Each atom was endowed with a
magnetic moment and ferromagnetism was the alignement of these
elementary entities occasioned by the existence of an external
magnetic field. The only trouble was that these fields were far too
small to maintain the alignement at room temperature, owing to
their thermal agitation. Weiss therefore proposed that there was an
internal magnetic field proportional to the magnetization
itself. Thus
\begin{equation}
\vec{H}_{Tot}=\vec{H}_{ext}+\vec{H}_{int}=
\vec{H}_{ext}+\alpha \vec{M}\, ,
\end{equation}
where M is the magnetization per atom.

The idea was matched neatly with van der Waals internal pressure in
the 20's when Heisenberg proposed his exchange mechanism wherein
there was an energy due to spin-spin interactions brought about by
interatomic coulombic forces (like those which were invoked to
explain chemical binding or Hund's rule for atoms). Thus Heisenberg
proposed an interaction
\begin{equation}
V_{spin-spin}=-v(r-r^{\prime})
\vec{S}(r).\vec{S}
(r^{\prime})\, ,
\end{equation}
$\vec{S}(r)$  being the spin of the atom located at $r$,
hence proportional to its magnetic moment. At typical interatomic
distances in a solid, $v(r)$ could be estimated to be something less
than chemical bond energies, rather like $O(10^{-1}ev)=10^{3}$ K.
The $v$ in Eq.(5) is not be confused with the $v$ of Eq.(1). It is the
spin-spin part. This advance lent more credence to Weiss's
suggestion since when the idea was first proposed,  the energy one
could come by was in dipole-dipole magnetic interactions and these
were too small by 3 orders of magnitude (Typically ferromagnetic
transitions occur at $O(10^{3}K)$~).

Thus Heisenberg proposed to furnish Weiss's hypothesis with a model
which contained spin-spin interactions in the form
\begin{equation}
V=-\frac{1}{2}\sum_{i,j}v_{ij}\vec{S}_{i}.
\vec{S}_{j}\, ,
\end{equation}
$\vec{S}_{i}$ being the spin on site $i$. This implies the
existence of an internal field given at site $i$ by
\begin{equation}
\vec{H}_{i}=\sum_{j}v_{ij}\vec{S}_{j}\, .
\end{equation}
MFA is then  the analog of van der Waals' approximation. One
neglects the correlation of $\vec{S}_j$ to $\vec{S}_i$
and approximates
\begin{equation}
\langle \vec{H}_{i}\rangle
=\vec{H} _{int}=\sum_{j}v_{ij}\langle
\vec{S}_{j}\rangle =\sum_{j}v_{ij}\langle
\vec M\rangle \, ,
\end{equation}
where we have used translational symmetry so that $\langle
\vec{S}_{j}\rangle $ is site independent. We have set the
elementary magneton of each atom equal to unity so that spin and
magnetic moment mean the same thing. Then $H$ has the dimensions of
energy.

Thus outfitted,Weiss's molecular field becomes (with $\widetilde{v}(q) =$
Fourier transform of $v_{ij}$)
\begin{equation}
\vec{H}_{mol}=\vec{H}+\widetilde{v}(o)
\vec{M}\, .
\end{equation}
From statistical mechanics one may then calculate $\langle
\vec{M} \rangle $ self consistently

\begin{equation}\langle \vec{M}\rangle =\frac{tr\exp
\left[ \beta 
\vec{H}_{mol}\cdot \vec{S}\right]
\vec{S}}{ tr\exp \left[ \beta
\vec{H}_{mol}\cdot
  \vec{S}\right] } \quad;\quad\beta =(1/kT)\, .
\end{equation}

\bigskip\bigskip\bigskip

{\bf \large II. Broken Discrete Symmetry}

\medskip
In the next few paragraphs we shall develop the idea using the
Ising model (proposed by Heisenberg to his student as a thesis
project). We shall see that this model is the prototype of a broken
discrete symmetry, as opposed to a continuous symmetry wherein
$S$ is a vector.

One treats $S$ as a 2-valued function, taking on values $\pm$1. Then
\begin{eqnarray}\langle M\rangle =\frac{\exp \beta
H_{mol}-\exp -\beta
H_{mol}}{\exp \beta
H_{mol}+\exp -\beta
H_{mol}}&=&\tanh\beta H_{mol} \nonumber\\&=&\tanh{\beta \left[ H+
\widetilde{v}(o)\langle M\rangle \right] }
\end{eqnarray}
a self consistent equation for $\langle M\rangle $. In Eq.(11) $H$
designates the external field. To see the consequences of Eq.(11) set
$H=0$. It is then seen that in addition to $\langle M \rangle =0$ two
additional solutions arise of equal and opposite values when $\beta
\widetilde{v}(o)>1$ since the slope of $\tanh x$ at $x=0$ exceeds unity
when $x>1$. These are the solutions which encode spontaneous
magnetization below the
 critical temperature ( $kT<kT_c$ where $kT_c=\widetilde{v}(o)$
). We shall
 shortly see that these are stable solutions whereas $\langle
M\rangle =0$ is unstable for $T<T_c $. This is SBS.

\hskip .5cm\epsfbox{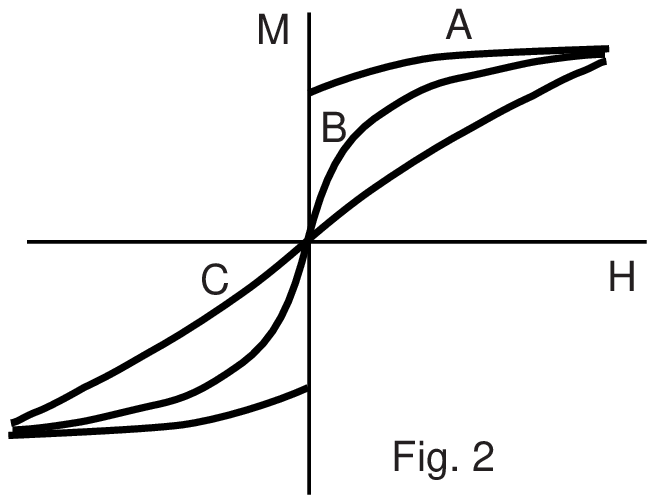}\hskip 1.5cm\epsfbox{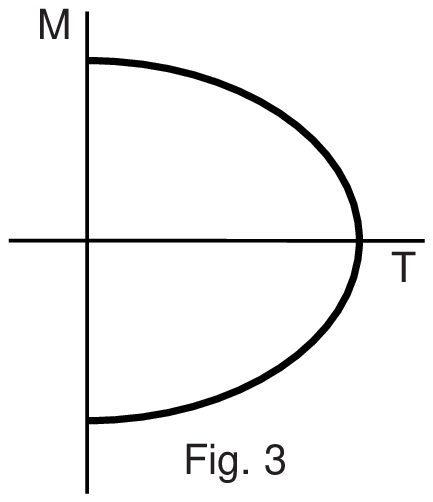}

In Fig.2, curve A is the isotherm for $T<T_c$, B for $T=T_c$ and C for
$T>T_c$.

The values of
$\langle M\rangle
$ at
$H=0$ are sketched in Fig.3.

Contact with the van der Waals theory is made as follows. Define
\begin{equation}
\epsilon _{i}=\frac{1}{2}(1+\mu _{i})\, .
\end{equation}
Thus $\epsilon =0$ means absence of a particle on a site and
$\epsilon =1$ means its presence. Then $\langle M\rangle $ is mean
density and $H$ is like pressure. Turning Fig.2 on its side and
tinkering with some thermodynamic identities converts Fig.2 into
Fig.1 . Fig.3 looked at on its side is the coexistence curve of
Fig.1 . In this case SBS is the choice of whether sites are occupied
by particles or holes. It has recently been proven by Fisher that
the analogy between liquid-vapor condensation in the case of
continuum space  (as opposed to the lattice) and ferromagnetism
runs very deeply, even to the most minute details of their critical
behaviour.

We shall continue with discrete SBS as exemplified by the Ising
model by deriving these results from an important construction
called the effective potential, a method developed by Bragg and
Williams in the early 30's (in another connection, but this is
irrelevant).

The partition function is
\begin{equation}
Z=tr\exp \left[ \frac{\beta }{2}\sum v_{ij}\mu
_{i}\mu _{j}+\beta
  \sum 
\mu_{i}H_{i}\right]\, .
 \end{equation}
For the nonce take $H_{i}$ independent of $i$. The trace is over all
$2^{N}$ spin states. Carry it out piecemeal $M$ by $M$ where $M$ now
means the total spin
\begin{equation}
M=\sum \mu _{i}=N_{up}-N_{down}\, ,
\end{equation}
so that
\begin{equation}
Z=\sum_{M}e^{\beta MH} tr_{M}\exp \left[ \frac{\beta
}{2}\sum
  v_{ij}\mu_{i}\mu _{j}\right] \, ,
\end{equation}
where $tr_M$ means summing over the $ {N \choose (N+M)/2}
$ states characterized by M.

Though one can carry through the construction in all rigor we shall
only develop the theory here in MFA in order to bring out the
essential ideas. It is the essence of MFA to neglect inter-spin
correlations.  Thus
\begin{eqnarray}
\frac{1}{2}\sum_{i,j}v_{ij}\langle \mu _{i}\mu _{j}\rangle _{MFA}&=
&\frac{1}{2}\sum_{i,j}v_{ij}\langle \mu _{i}\rangle \langle \mu
_{j}\rangle \\  &=&\frac{1}{2}\sum_{i,j}v_{ij}\langle \mu
\rangle _{M}^{2}\, ,
\end{eqnarray}
where we used translational symmetry to set $\langle \mu
_{i}\rangle $ independent of site $i$. The symbol $\langle \mu
\rangle _{M}$ means the average of a spin in the subensemble
characterized by M i.e.
\begin{equation}\langle \mu \rangle _{M}=\frac{1}{N}\sum_{i}\langle
\mu _{i}\rangle _{M}=\frac{M}{N}={\it m}\, .
\end{equation}
Thus in MFA one has
\begin{eqnarray}Energy &=&-\frac{1}{2}\sum_{i,j}v_{ij}\mu _{i}\mu
_{j}-H\sum \mu _{i}\nonumber\\ &=&-\frac{1}{2} N \widetilde v(0) m^2 -NHm\ .
\end{eqnarray}
Thus
\begin{equation}
Z_{M}=\Omega (M)e^{-\beta E(M)}\, ,
\end{equation}
where
\begin{equation}
\Omega (M)= {N \choose (N+M)/2}\, ,
\end{equation}
whence
\begin{eqnarray}
\ln Z_{M}&=&\ln \Omega (M)-\beta E(M)\nonumber\\
\ln \Omega (M)&=&N\ln 2-\frac{N}{2}\left[ \left( 1+m\right) \ln
(1+m)+(1-m)\ln (1-m)\right] , 
\end{eqnarray}
where we have used Stirling's approximation and Eq (18). $\ln Z_{M}$
has a sharp maximum at $N=Mm^*$ where
\begin{equation}
m^{\ast }=\tanh \left[ \beta
\widetilde{v}(o)m^{\ast }+H\right]\, ,
 \end{equation}
the relative width of which is $O(1/\sqrt N)$ so that in the
thermodynamic limit (i.e. $\lim_{N\to \infty }\ln Z$\ ) one
has
\begin{equation}
\ln Z=\ln (Nm^{\ast })\, .
\end{equation}
Since $\ln Z=-\beta$[{\it Helmholtz free energy}\/], we
identify $\ln
\Omega (Nm^{\ast })$ with the entropy (because the energy has
already been identified in Eq (19) ). Over the years we have come to
call $(-1/N)\ln Z $ the effective potential and this has become the
standard way to approach SBS in field theory (since $Z(M)$ is the
functional integral over configurations of $\exp\, (-S)$ where $S$ is the
eucledeanized action; in our case the functional integral is the discrete
sum over $2^{N}$ configurations).

One gets a first glimpse into the field formulation by looking at
$V_{eff}$ for small~$m$
\begin{eqnarray}V_{eff}=-\lim (\frac{1}{N})\ln
Z_{N}&=&-\frac{1}{2}\beta \widetilde{v} (o)m^{2}-\beta
mH+\frac{1}{2}m^{2}+\frac{1}{12}m^{4}+\ldots\nonumber\\ &=&\frac{1}{2}
(1-\beta
\widetilde{v}(o))m^{2}+\frac{1}{12}m^{4}-\beta mH\, ,\end{eqnarray}
$m$ is to be considered a field taking on a continuum of values in
the $N\to\infty$ limit and from now on we shall use the
symbols $m$ and $\varphi$ (for field ) interchangeably. In Eq.(25) the
irrelevant constant $N\ln2$ has been dropped.

From Eq.(25) we have,
\begin{equation}
V_{eff}=\frac{1}{2}\mu ^{2}\varphi ^{2}+\lambda
\varphi^{4}-\varphi H\, ,
\end{equation}
where $\mu ^{2}=(1-\beta \widetilde{v}(o))=(1-(T_{c}/T))$. It is seen
that  $\mu^{2}$ changes sign at $T=T_{c}$ , becoming negative for
$T<T_{c}$. $V_{eff}$ is sketched in Fig.4

\hskip .5cm\epsfbox{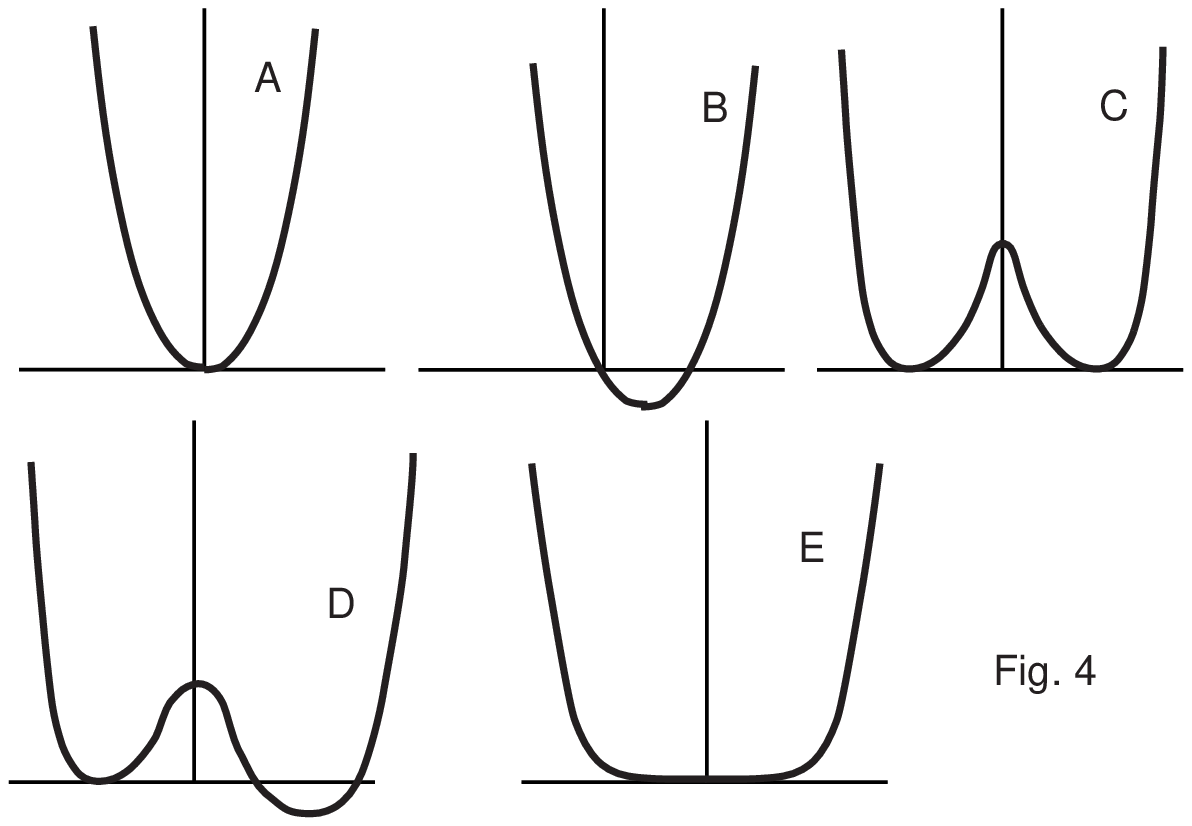}

\begin{quote}
\small
A: $T>T_{c}\, ;H=0$ ~~~B: $T>T_{c}\, ;H>0$ ~~~C: $T<T_{c}\, ;H=0$ ~~~D:
$T<T_{c}\, ;H>0$ ~~~E:~$T=T_{c}\, ;H=0$

These pictures essentially contain the whole story and nothing more.
\end{quote}

\normalsize
For $2<d\leq 4$, small modifications exist in the critical region
defined by $\left\vert 1-\frac{T_{c}}{T}\right\vert =O\ $(a few
percent) and $H/kT$ a few percent. Otherwise MFA has quantitative
significance, and always qualitative significance. For example for
$d=3$ one has $\mu ^{2}\sim
\left\vert T-T_{c}\right\vert ^{0.62}$  in the critical region at
$H=0$ rather than $\mu ^{2}\sim \left\vert T-T_{c}\right\vert ^{0.5}$. The
latter estimate becomes  valid once $\left\vert
1-\frac{T_{c}}{T}\right\vert>10\%$.  The next section will afford further
insight into the whys and wherefores of these facts.
\bigskip

{\bf \large III. Correlation Function (Green's Function)}
\medskip

Consider $H=0$ and $T=T_{c}+\epsilon $ where $\epsilon $\ is small
and positive. As $\epsilon \rightarrow 0$, long range order comes
into being i.e. if one fixes the orientation of a single spin out
of $N$, then all $N$ get oriented in the same direction. It then must
be expected that for $\epsilon >0$, there must be a precursor of
this phenomenom i.e. long range order should be heralded in by
correlations among spins over increasingly longer range as
$\epsilon \rightarrow 0$. The theory of this was worked out in the
first decades of the 20th century by Ornstein-Zernike, by
Smolochowski and by Einstein, the interest being the tremendous
enhancement of light scattering by critical fluctuations (i.e. long
range correlations) giving rise to critical opalescence. In our
case the cross section is proportional to $
\langle \left\vert \mu _{q}\right\vert ^{2}\rangle $ where
\begin{equation}
\mu_{\vec q}=\sum_{i}\mu
_{i}\exp \left[ i
\vec{q}\cdot \vec{R}_{i}\right]\, ,
 \end{equation}
hence to the Fourier transform $\langle \mu _{i}\mu _{j}\rangle $,
therefore at small $q$ sensitive to long range correlations.

A first shot at the problem is contained in a rather obvious
generalization of MFA. Turn on an external field which varies from
site to site
\begin{eqnarray}
Hamiltonian=\cal H &=&-\frac{1}{2}\sum v_{ij}\mu _{i}\mu
_{j}-\sum \mu _{i}H_{i}\nonumber\\ Z&=&tre^{-\beta \cal H}\, .
\end{eqnarray}
Then
 \begin{equation}
\langle \mu _{i}\rangle =\frac{\partial
\ln Z}{\partial \beta H_{i} }\, ,
\end{equation}
\begin{equation}\langle \mu _{i}\mu _{j}\rangle -\langle
\mu _{i}\rangle \langle
\mu _{j}\rangle =\frac{\partial \langle \mu _{i}\rangle }{\partial
\beta H_{j}}=\frac{\partial ^{2}\ln Z}{\partial \beta H_{i}\partial
\beta H_{j}}\, .
\end{equation}
For $T>T_{c}$ and in $\lim_{H_{i}\rightarrow 0}$, all $i$, Eq.(30)
allows to compute the (connected) correlation function (Green's
function in field theory). We shall approximate this by use of MFA
 in this extended local sense.

The field on $\mu _{i}$ is $\sum_{j}v_{ij}\mu_j+H_{i}$. Thus for small
$H_{i}$ the average of $\mu _{i}$ for $T>T_{c}$ is
\begin{equation}
\langle \mu _{i}\rangle =\tanh \beta (\sum
v_{ij}\langle \mu _{j}\rangle +H_{i})\rightarrow \beta \left[\sum 
v_{ij}\langle \mu _{j}\rangle +H_{i}\right]\, . 
\end{equation}

The essential approximation that has been made is that $\langle \mu
_{j}\rangle $ is calculated with a probability distribution that is
independent of the orientation of $\ \mu _{i}$~. This is not exact
since the distributions differ according to $\mu _{i}=+1$ or $\mu
_{i}=-1$. This neglect of correlation in the present context is
then a sort of local MFA. Taking a derivative of Eq.(31) with respect
of to $H_k$ and using Eq (30) gives
\begin{equation}
\langle \mu _{i}\mu _{k}\rangle =\delta
_{ik}+\beta \sum v_{ij}\langle \mu _{j}\mu _{k}\rangle\, .
\end{equation}
This is like an integral equation for $G_{ij}(\equiv \langle \mu
_{i}\mu _{j}\rangle )$. It may be solved by Fourier transform. Denoting by
$G(q)$ and $\widetilde{v}(q)G(q)$ the Fourier transforms of $G_{ij}$ and
$v_{ij}$, one has
\begin{equation}
G(q)=1+\beta
\widetilde{v}(q)G(q)\, ,
\end{equation}
\begin{equation}G(q)=\frac{1}{1-\beta
\widetilde{v}(q)}\, .
\end{equation}

An interaction which is ferromagnetic over its whole range has
$v(\vec {R }_{i}-\vec {R}_{j})>0$ for all distances
$\left\vert \vec{R}_{i}-
\vec{R}_{j}\right\vert $. Therefore $\widetilde v(q)$ is maximal at $q=0$
and has the form
 \begin{equation}
\widetilde{v}(q)=\widetilde{v}(0)-\alpha q^{2}\, ,
\end{equation}
valid small $q$ ($qa\ll 1$ where $a$ = lattice distance).

Thus
\begin{equation}
G(q)=\frac{1}{(1-\beta \widetilde v(0))+\alpha
q^{2}}\simeq \frac{1}{\mu ^{2}+\alpha q^{2}}\, ,
\end{equation}
and we see that the curvature of the effective potential at its
minimum (for 
$T>T_{c}$\ and $H=0$ ) is equal to the $(mass)^{2}$ in $G(q)$. In
this way one sees that for small values of $\left[(T-T_{c})/T_{c}\right] $
with 
$ (T-T_{c})>0$ and for small values
of $\left( H/kT_{c}\right) $ the spin system is governed by an
effective action density equal to
\begin{equation}\frac{1}{2}(\nabla \varphi )^{2}+(\mu
^{2}/2)\varphi ^{2}+\lambda
\varphi ^{4}+\varphi H\, ,
\end{equation}
with $\mu ^{2}\sim (T-T_{c})$ and $\lambda >0$. We have dropped
irrelevant factors of $O(1)$ which may be absorbed into the
definition of $\varphi $, $
\mu $ and $\lambda $. The important point is that $\mu
^{2}\rightarrow 0$\ as $T\rightarrow T_{c}$ and one is confronted
with an infra red problem at $ T=T_{c}$ , $H=0$. This gives rise
to the theory of critical phenomena which results in a dynamical
theory of renormalization. In particular mass renormalization
shifts $\mu ^{2}$ to $\sim (T-T_{c})^{0.62}$ for $d=3$. This is of
little interest to us in this review which is an exploration of the
physical mechanism behind SBS. Nevertheless it is interesting to
understand how it is that there is a threshold value of $d$\, $(d>4)$ 
for which these renormalization effects become insignificant.

From Eq.(32), one sees that $G(R_{i}-R_{j})$ is built out of chains
of the interactions $v_{kl}$ (Fig.5).

 \hskip  2cm\epsfbox{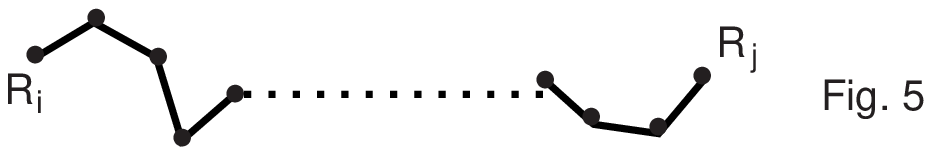}

The sum of all these random walks is $G_{ij}$. One can recast these
considerations so as to take into account  the fact that in the
correct rule for each walk, a given intermediate spin is visited
one and only one time. To this must be added walks which do have
intersections. These have different weights. One can then convert
the problem into a random walk without restriction giving rise to our
$G(q) $  plus corrections due to intersections (there has been some
oversimplification but without injustice to the essential physics). It is a
fact, (and one can show it from the field theory itself) that for
$d>4$, the probability of self intersection becomes so small that
it has no effect on the Green's functions. There is too much space
around and that is why MFA becomes exact (except for irrelevant
factors of O(1) ) for $d >4$.

Now let us see what happens \ $T<T_{c}$. Then $\langle \mu
_{i}\rangle =m$ and the Green's function is
\begin{equation}
G_{ij}=\langle \mu _{i}\mu _{j}\rangle -\langle \mu
_{i}\rangle \langle \mu _{j}\rangle =\langle \mu _{i}\mu
_{j}\rangle -m^{2}\, .
\end{equation}

One can go through some formalism to establish the rules for how to
construct the random walk in this case. Suffice it to say that,
once again, one sums on random walks in MFA, but one must weight
each vertex with the factor $(1-m^{2})$. To understand this
physically, note that when  one asks for the infinitesimal variation of
$\langle
\mu _{i}\rangle $
due to a variation of $H_{i}$ , the field on $\mu _{i}$ , then
\begin{equation}
\frac{\partial \langle \mu _{i}\rangle
}{\partial \beta H_{i}}=
\frac{\partial }{\partial \beta H_{i}}\left[ \tanh \beta
H_{i}\right] _{H_{i}=H_{i}^{0}} = 1-m^2(H_i^0)\, .
\end{equation}
Here $H_{i}^0$ is the value of the field on $\mu _{i}$ before the
variation, be it an external or internal field!

Since in the chain of interactions, an intermediate spin, say $k$ , is
submitted to a variation of the field $H_{k}$ upon it, coming from
the link which precedes it, (where one considers the chain
originating at $i$ and terminating at $j$). The response of $\ \mu
_{k}$ \ to this variation is equal to $\beta v_{lk}(1-m^{2})$ which
is taken to be small.

The net result is for $T<T_{c}$ , $H=0$
\begin{equation}G(q)=\frac{1-m^{2}}{1-(1-m^{2})\beta
\widetilde{v}(q)}\, ,
\end{equation}
which for small q reads
 \begin{equation}
G(q)=\frac{1}{\mu ^{\prime 2}-\alpha
q^{2}}\, ,
\end{equation}
where
\begin{equation}
\mu ^{\prime
2}=\frac{1}{(1-m^{2})^{-1}-\beta \widetilde{v}(0)}\, .
\end{equation}
In this expression $m$ is the spontaneous magnetization i.e.
\begin{equation}m=\tanh \left[ \beta
\widetilde{v}(0)m\right]\, . \end{equation}
From Eq.(43) and Eq.(42) it is very easy to show that $\mu ^{\prime
}{}^{2}>0$ and in fact for small $(T_{c}-T)/T_{c}$ one has $\mu
^{\prime }{}^{2}\sim (T_{c}-T)$.

Thus the wild infra-red fluctuations encountered as $T\rightarrow
T_{c}+\epsilon $\ with $\epsilon >0$ become quenched through the
existence of m for $T<T_{c}$ . One easily shows that $\mu ^{\prime
}{}^{2}$ is the curvature of $V_{eff}$ in Fig.4C at one of the
minima.

One defines the susceptibility, $\chi $ , as $(\partial
m/ \partial H)\vert _{H=0}$ and in both cases (\ $(T-T_{c})$\ positive
or negative) one has $
\chi \sim \left\vert T-T_{c}\right\vert ^{-1}$ .
\bigskip

{\bf \large IV. Broken Continuous Symmetry}
\medskip

Rather than the spin taking on discrete values like $\mu _{i}=\pm
1$, one can now study a spin which is a vector. This can be done
classically by placing a unit vector on each lattice site. Then the
trace that is used to calculate $Z$ is
\begin{equation}
h=\int 
\mathop{\prod} d\vec{S}_{i}~ \delta (\sum
\vec{S}_{i}^{2}-1)\, ,
\end{equation}
where $\vec{S}_{i}$ is an $n$ dimensional vector in some
internal space. Or one can do a quantum calculation where
$\vec{S}_{i}$ is an operator and the trace is the sum
over the eigenvalues of $
\vec{S}_{i}$ in some group representation. For example if
$
\vec{S}_{i}=\vec{\sigma }_{i}$\ \ (=Pauli
matrices) then the trace is once again over the values $\pm $1 for
each spin. The Hamiltonian in either case is taken symmetric with
respect to the transformation of the symmetry group which is
represented by $
\vec{S}_{i}$
\begin{equation}
{\cal H}=-\frac{1}{2}\sum
v_{ij}(\vec{S}_{i}\cdot 
\vec{S}_{j})\, ,
\end{equation}
and one represents an external breaking by fields $H_{i}$ according
to
\begin{equation}{\cal H}_{ext}=-\sum \vec{S}_{i}\cdot
\vec{H}_{i}\, .
\end{equation}

For \ $T>T_{c}$\ the physics of this case resembles strongly that
of the Ising model. For example if
$\vec{S}_{i}=\vec{\sigma } _{i}$\ \ then the
chain of interactions contributing to $G(q)$ contains chains like
\begin{equation}
 tr\left[ \ldots v_{kl}\vec{\sigma
}_{k}.\vec{
\sigma }_{l}~v_{lm}\vec{\sigma
}_{l}.\vec{\sigma } _{m}\ldots \ \right]\, .
 \end{equation}

Since $tr\vec{\sigma }_{l}\vec{\sigma }_{l}$
is $\neq 0 
$ only for equal spatial components, one sees that the trace is the
same as that for the Ising model. Then the chain which contributes
to $\left\langle
\sigma _{i}^{x}\sigma _{j}^{x}\right\rangle $ is the same as for
the Ising model. Moreover $\left\langle \sigma _{i}^{x}\sigma
_{j}^{y}\right\rangle =0$. Thus once again,in MFA one has
\begin{equation}G(q)\sim \frac{1}{q^{2}+\mu
^{2}}\, .
\end{equation}

However for $T<T_{c}$ , new physics emerges. The easiest road to
Rome passes by the effective potential. From Eq.(26), one sees that the
system duplicates in detail the theory of discrete SBS.  Thus the
quadratic part of the potential is proportional to $(\nabla 
\vec{\varphi })^{2}+\mu ^{2}\vec{\varphi
}^{2}$ where $
\mu ^{2}=T-T_{c}$. As before, there will be corrections due to self
intersections of walks (and other dynamic effects which in fact are
dependent on the representation $\vec{\varphi }$ of the
symmetry group in question. But, whatever, the form of the quartic
interaction is dictated by symmetry to be $\lambda \left[ \left(
\vec{\varphi } _{i}\right) ^{2}\right] ^{2}$). [This is
going a little too fast since sometimes other polynomial forms are
available, (like $d_{ijk}\varphi _{i}\varphi _{j}\varphi _{k}$\
for $SU_{3}$ with $\varphi _{i}$ in the regular representation),
but this cursory review is not the place to enter into such
niceties]. Thus one is led to
 \begin{equation}V_{eff}=\frac{1}{2}\left[ \nabla
\vec{\varphi
    }^{2}+\mu^{2}(\vec{\varphi })^{2}\right] +\lambda
(\vec{\varphi   }^{2})^{2}+\vec{\varphi
}\cdot \vec{H}\, .
\end{equation}

Whereas $\mu ^{2}$ changes sign at $T=T_{c}$ ( being proportional
to $ (1-\beta \widetilde{v}(0))$\ up to a scaling of
$\widetilde{v}(0)$ ), one has $\lambda >0$. This is most easily
seen in MFA where $\lambda $ arise from the entropy factor as in
the Ising model.

It then follows that all the pictures of Fig.4 remain applicable
for use of SBS in the case of continuous symmetry provided they
become multidimensional in ``$\varphi $-space''. For ease of
representation let the symmetry be $U(1)$. Then
$\vec{\varphi }=(\varphi _{1},\varphi _{2})$ and the
pictures must be interpreted as planes cut through figures of
revolution about a central axis in $\varphi $ space.

The dimple in Fig.4C is the unstable solution
$\vec{\varphi }=0$ for $T<T_{c}$. The more stable
minima lie along a circle in the $\varphi _{1},\varphi _{2}$ plane
$\vec{\varphi }^{2}=\varphi _{1}^{2}+\varphi
_{2}^{2}=m^{2}$ where $\partial V/\partial (\vec\varphi ^{\,2})=0$. In
this case SBS is the choice of which vector
$\vec{\varphi }$ is taken non vanishing. The reason for
the words ``more stable'' rather than ``stable'' is clear. Suppose
$\vec{H}\neq 0$\ in some direction then the minimum
becomes stable and $\vec{\varphi }\parallel 
\vec{H}$. Now let $H\rightarrow 0$. The solution then
tends to that value of $\vec{\varphi }$ which is on the abovementioned
circle without changing its direction as one takes the limit. But it is
unstable with respect to directional changes upon applying an
infinitesimal adjunction of
\ $\vec{H} ^{\prime }$ in a direction different from the
original $\vec{H}$ (which had been sent to zero). This will
cause $\vec{\varphi }$ to swivel along the circle so as
to lie in the direction of
$\vec{H}^{\prime }$, albeit this latter is
infinitesimal but not zero.

\bigskip

This new element of SBS of continuous symmetry is essential to the
physics of all kinds of situations and as will be seen in the gauge
section, plays a vital role in the Brout-Englert-Higgs (BEH)
mechanism\footnote{References to the gauge theory and relevant
material are given in ``Modern Times''. }. It is an effect which was first
consciously put on display by Felix Bloch in the mid 1930's in his spin
wave theory of ferromagnetism. In the next sections we shall review his
theory, as well as applications to superconductivity and superfluidity.
This will be followed by Nambu's development of the theory of
spontaneously broken chiral symetry and soft pion physics. The
expression of these ideas in terms of a relativistic field theory, often
called the Goldstone theorem, will be presented in the context of the
BEH mechanism in the section ``Modern Times'' by Fran\c{c}ois Englert
since it is herein that this aspect of the theory has been particularly
successful.

We briefly summarize the important result of SBS of continuous
symmetry which has been deduced up to this point. In terms of the
effective potential for $ V_{eff}(\vec{\varphi })$ , for
$T>T_{c}$ the situation is the same as for discrete symmetry. There
is a unique minimum at $\vec{
\varphi }^{2}=0$, whose curvature is the inverse susceptibility
$\chi =
\partial m/\partial H=(\mu ^{2})^{-1}\sim (T-Tc)^{-1}$.
For $ T<T_{c} $, the point $\vec{\varphi }^{2}=0$ is a
local maximum since at this point the curvature ($=T-T_{c}$) is
then negative. There is then an ``orbit'' of minima which for the case of
broken $U(1)$ symmetry is a circle (for the general case see Modern
Times for a description of this orbit in the space of a
representation of a general group). From our discussion it is seen
that the susceptibility becomes a tensor in ``$\vec{\varphi }
$'' space. One defines a longitudinal susceptibility corresponding
to the response of $\left\langle \vec{\varphi
}\right\rangle $\ with respect to $\vec{H}$ parallel to
an priori fixed vector $
\vec{\varphi }_{0}$ which is on  the orbit of minima at $H=0$.
One thinks of this as a ``stretching'' mode of response of the
magnetization. The transverse susceptibility is the response to
$\vec{H}$ orthogonal to $\left\langle
\vec{\varphi }\right\rangle _{0}$ and according to our
discussion of instability this is infinity. One defines a
$(mass)^{2}$ tensor which is $\chi ^{-1}$ whereupon $\mu
_{longitudinal}^{2}\sim (T_{c}-T)$
and $\mu _{transverse}^{2}=0$. This vanishing of the mass in the
transverse direction is in fact the terminal point of a continuous
spectrum of excitations, the modes being sorted out according to
Fourier transform. The expression of this in relativistic field
theory (Goldstone's Theorem) is covered in ``Modern Times". The
application to feromagnetism is very instructive in this regard.
This will be the subject of the next section.

\bigskip
{\bf \large V. Spin Wave Theory}
\medskip

The existence of zero mass modes as collective excitations (i.e.
bosons in quantum field theory) is neatly revealed in spin wave
theory. We here follow a procedure, due to Bloch, by studying the
single quantum excitations from the the ground state (vacuum). For
simplicity we work with $\vec{S
}_{i}=\vec{\sigma }_{i}$ (the Pauli matrices). SBS is
the choice of orientation of vacuum, the only group symmetric
specification for which is ``all spins parallel'' i.e. all in the
same spin state. We choose $\mid 0>$ to the ``all spins up'' i.e.
$\sigma _{i}^{z}\mid 0>=$ $\mid 0>$.

Excitations then are generated by creating down spins, the most low
lying being l.c.'s of $\sigma _{i}^{-}\mid 0>$. These l.c.'s are
determined from
\begin{equation}
\left[ H,\varphi _{\omega }\right] \mid
0>=i\omega \varphi _{\omega }\mid 0>\, ,
\end{equation}
where $\varphi _{\omega }$ is an ``eigen operator'', i.e an l.c. of 
$\sigma_{i}^{-}\mid 0>$ which satisfies $\left[ H,\varphi _{\omega }\right]
= i\omega \varphi _{\omega }$. Using the algebra of Pauli matrices, along 
with $\left[
\vec{\sigma }_{i},\vec{\sigma }_{j}\right] =0$ for
$i\neq j$, we get with ${\cal H}=-\frac{1}{2}\sum v_{ij}\vec{
\sigma }_{i}\cdot \vec{\sigma }_{j}$,
\begin{equation}\left[ {\cal H},\sigma _{i}^{-}\right] =-\sum
v_{ij}(\left[ \sigma _{i}^{z},\sigma _{i}^{-}\right] \sigma
_{j}^{z}+\left[ \sigma _{i}^{+},\sigma _{i}^{-}\right] \sigma
_{j}^{-})=\sum_{j}v_{ij}(\sigma _{j}^{z}\sigma _{i}^{-}-\sigma
_{i}^{z}\sigma _{j}^{-})\, .
\end{equation}

Operating on vacuum, we
get
 \begin{equation}\left[ {\cal H},\sigma _{i}^{-}\right] \mid
0>=\sum_{j}v_{ij}(
\sigma_{i}^{-}-\sigma _{j}^{-})\, .
\end{equation} 

By translational symmetry Eq.(52) is diagonalized by Fourier
transform. Defining $\sigma _{q}^{-}=(1/\sqrt{N})\sum \sigma
_{i}^{-}e^{i
\vec{q}\cdot \vec{R}_{i}}$ one has
 \begin{equation}\left[ {\cal H},\sigma _{q}^{-}\right] \mid
0>=i\omega _{q}\sigma _{q}^{-}\mid 0>=\left[
\widetilde{v}(0)-\widetilde{v}(q)\right] \sigma _{q}^{-}\mid
0>\, ,
\end{equation}
where
\begin{equation}\omega
_{q}=\widetilde{v}(0)-\widetilde{v}(q)\sim q^{2}
\quad , \quad \hbox{small}~ q\, . 
\end{equation} 

The generalization to the case of an external field is equally
interesting. Clearly our vacuum $\mid 0>$, corresponds to $\vec H$ in the
$z$ direction. So adding to $\cal H$ a term $-H\sum \sigma _{i}^{z}$, going
through the same steps then leads to
\begin{equation}\omega
_{q}=\widetilde{v}(0)-\widetilde{v}(q)+{\it H}\sim q^{2}+ {\it
H}\, .
\end{equation}

Thus $H$ induces a $(mass)^{2}$ in the zero mode which is
linear in the external breaking. This is especially important when
applying these ideas to SB$\chi$S and soft pion physics. It is to be
noted that the excitation operator, $\sigma _{q}^{-}$, reduces to the
global rotation operator at $q=0$ i.e. $\left[ H,\sigma _{q=0}^{-}\right]
=0$ (at $H=0$) in virtue of symmetry whence $\omega (0)=0$, and we
see that $\mu _{transverse}^{2}=0$ is indeed the statement that the
excitation energy of a continuous spectrum vanishes at $q=0$ in virtue of
symmetry.

We also can now see why continuous SBS cannot apply in its
na\"{\i}ve form to $d=2$. The number of spin waves, at temperature
$\beta ^{-1}$, is $\left[ e^{\beta \omega _{q}}-1\right] ^{-1}$\
for $kT\ll \widetilde{v}(0)$ (for higher T they interact and the
ideal gas of excitations is no a longer valid approximation). Then
the total number of spin waves at low $T$ is
\begin{equation}\sim \int d^{d}q\frac{1}{e^{\beta \omega
_{q}}-1}\sim \beta ^{-1}\int \frac{d^{d}q}{q^{2}}\, ,
\end{equation}
which diverges in the infra-red at $d=2$ i.e. $\mid 0>$ is unstable
for $H= 0$. New methods are therefore required in
continuous SBS. But for SBS in the discrete case, the naive notions
are OK, albeit suffering severe quantitative modifications.

Some conceptual issues arise which we will now address. Their resolution
is of pedagogical interest especially when compared with the
corresponding situation in the gauge theory.

We shall first display the classical concept of broken symmetry given by
the familiar picture of an arrow which points in the ``direction of the
vacuum state'' picked by the broken symmetry.  For example, in the
above paragraphs this arrow points in the $z$-direction of group space.
For simplicity, we continue with the example of broken $SU(2)$
symmetry represented by a Pauli matrix sitting on each lattice site,
wherein the Hamiltonian is a group scalar as in Eq.(45). The
generalization of these considerations to any group in any
representation is straightforward. 

Let $\vert 0\rangle$ be the vacuum state: $S_z  \vert 0\rangle = N/2
\vert 0\rangle $ where
\begin{equation}
\vec S = \sum_i {\vec \sigma_i\over 2} \, .
\end{equation}
Since the $S_\alpha (\alpha = x,y,z)$ represent group generators (i.e. $[
S_\alpha, S_\beta] =i \epsilon_{\alpha\beta\gamma} S_\gamma )$, one
may construct a rotated vacuum from them. For example, a  rotation
about the $x$-axis of $\vert 0\rangle$ gives the rotated vacuum  $\vert
\theta\rangle$ where
\begin{equation}
\vert \theta\rangle = e^{iS_x\theta} \vert 0\rangle\, .
\end{equation}
The states  $\vert
\theta\rangle$ and $\vert
0\rangle$ are degenerate since $[{\cal H}, S_x] =0  $,  ${\cal H}$ being
scalar and $\vec S$ being a group vector.

Since $S_x$ is a group generator, it  follows that
 \begin{equation}
\langle\theta\vert S_x\vert\theta\rangle=0 ~;~\langle\theta\vert
S_y\vert\theta\rangle=-\sin\theta~;~\langle\theta\vert
S_z\vert\theta\rangle=\cos\theta\, .
\end{equation}
In this way, the classical notion of ``arrow'' is given by the expectation
value of the operator $\vec S$ in the different rotated vacua.

We shall now prove that, for $\theta$ fixed, in the limit $N\to\infty$,
$\langle\theta\vert0\rangle =0$. Moreover, we shall show that the
Hilbert space of excitations built upon different vacua are mutually
exclusive as well (in the limit $N\to\infty$).
\begin{eqnarray}
\langle 0\vert\theta\rangle &=&\langle 0\vert e^{iS_x\theta} \vert
0\rangle = \langle 0\vert \prod_{i=1}^N   e^{i(\sigma_i^x /2)\theta} \vert
0\rangle\nonumber\\
&=&\prod_{i=1}^N  \langle 0\vert \cos\,  (\theta/2)  +
 i (\sigma_i^x /2)\sin\, (\theta/2)\vert
0\rangle\nonumber\\
&=& [ \cos\,  (\theta/2)]^N \quad \longrightarrow_{N\to\infty}~ 0\, .
\end{eqnarray}

If instead of the overlap of $\langle 0\vert$ with $\vert\theta\rangle$
we took excited states of   $\langle 0\vert$, say containing $n$ spin
wawes, the overlap would then be $\sim  [ \cos\,  (\theta/2)]^{N-n}$. So
even if $n$ is a finite fraction of $N$, the result vanishes in the limit.
With more effort one can prove that the excited states built on
$\vert\theta\rangle$ are orthogonal to excited states of 
$\vert 0\rangle$. This remains true until one reaches some
threshhold number of excitations proportional to $N$ at which point one
approaches critical conditions wherein these naive considerations break
down.

For finite $N$ one can always construct $N+1$ orthogonal ``vacuum''
states  as one does in the conventional method of quantizing angular
momenta. These are the states $(S^-)^p \vert 0\rangle\,  ;\,  p=0,1 ... \, ,
N$. States corresponding to a rotation $\theta, \varphi $ from
$\vert 0\rangle$ are obtainable as a linear combination of these. For
finite $N$ such states are not, in general,  orthogonal. But they become
approximately so when their angular difference exceeds $O(1/\sqrt N)$.
In this way one recovers their mutual orthogonality as $N\to\infty$ for
any  angular difference.

\bigskip

{\bf \large VI. Superfluidity and Superconductivity}
\medskip

We briefly indicate how SBS applies to these two interesting
phenomena. A free boson gas of $N$ particles condenses at a temperature
for which the thermal Compton wave length $(mkT)^{-1/2}$ is
$O$(interparticle distance). For $T<T_{c}$ , a finite fraction of $N$
occupies the state $k=0$, and at $T=0$ all $N$ have zero momenta. For
the interacting case, at $T=0$ there is only a finite fraction which
condenses i.e.
\begin{equation}
\left\langle a_{0}^{+}a_{0}\right\rangle
=N_{0}=\alpha N \quad ,
\quad \alpha <1\, .
\end{equation}

   This macroscopic occupation of the $k=0$ state can be transcribed
into a SBS as follows. The commutator $\left[ a_{0}^{+},a_{0}\right]
=1$ is negligible with respect to $N_0$ i.e. $N_{0}\cong N_{0}+1$ in
good approximation. Then one can treat $a_{0}$ as a c-number. But
$a_{0}$ has a phase. The choice of this phase is SBS. Bogoljubov~\cite{bog}
built a system of excitations in analogy to spins waves, by building
them from a vacuum with a fixed complex c-number value of $ a_{0}$. They
are linear combinations of the form $\Psi _{q}^{+}=\alpha
_{q}a_{q}^{+}+\beta _{q}a_{-q}$ (note $a_{-q}$ $\mid 0>\neq 0$ because $\mid
0>$ contains virtual occupation of states with $q\neq 0$ , in virtue of the
interatomic interactions). The Bogoljubov coefficients $\alpha _{q},\beta
_{q}$ (with ($\left\vert \alpha _{q}\right\vert ^{2}-\left\vert \beta
_{q}\right\vert ^{2})=1$) are proportional to $a_{0}$ and $a_{0}^{\ast }$
respectively. The point to be made here is that as $q\rightarrow 0$, the
operator
$\Psi _{q}^{+}$ becomes a rotational generator in the ``gauge plane''
i.e. it generates infinitesimal changes of the phase of $a_{0}$.

Superfluidity is then a spectacular example of SBS where the
symmetry is $U(1)$. The all important phase plays vital physical role
since if one lets it vary from point to point, its gradient is the
velocity of superfluid. 

Superconductivity is an equally fascinating
case of spontaneously broken $U(1)$ symmetry. Bound states (Cooper
pairs) are s states in spin singlets, so causing correlations
$\left\langle n_{k\uparrow }n_{-k\downarrow }\right\rangle
-\left\langle n_{k\uparrow }\right\rangle \left\langle
n_{-k\downarrow }\right\rangle $\ which are $O(1)$ rather than the
usual free gas value $O(1/N)$. In terms of a pseudo spin algebra
which is isomorphic to Pauli spin matrices given by
\begin{eqnarray} b_{k}=a_{k\uparrow }a_{-k\downarrow }&\sim& \sigma
_{k}^{-}\nonumber\\ b_{k}^{+}=a_{-k\uparrow }^{+}a_{k\downarrow
}^{+}&\sim& \sigma _{k}^{+}\nonumber\\  1-n_{k\uparrow
}-n_{-k\downarrow }\ &\sim& \sigma _{k}^{z}\, ,
\end{eqnarray}
one invents a set of order parameters which are $\left\langle
b_{k}\right\rangle $. Since $b_{k}$ has a phase, one breaks $U(1)$
and since the hamiltonian is invariant under this $U(1)$ symmetry,
the interactions being $v(k,k^{\prime })b_{k}^{+}b_{k^{\prime }}$, 
one has SBS. Note that an otherwise $SU(2)$ symmetry is broken
externally since the kinetic energy in the Hamiltonian is equal to
$\sum \varepsilon (k)(n_{k\uparrow }+n_{-k\downarrow })$ hence up
to a constant $=\sum \varepsilon (k)\sigma _{k}^{z}$.

A typical ``vacuum'' configuration may be depicted as follows as one
spans the Fermi surface in $k$-space

\hskip 2cm\epsfbox{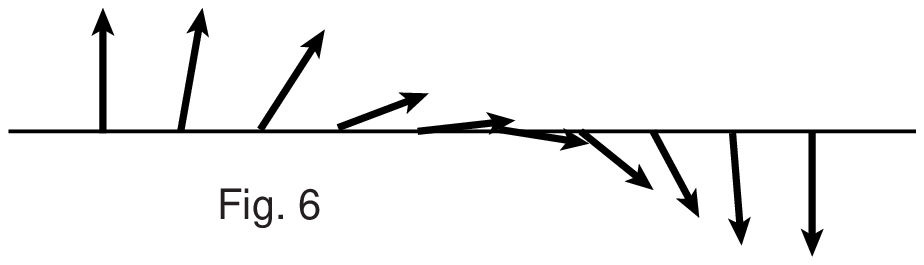}

(in the free or normal metal at zero temperature $\left\vert
k\right\vert =k_{f}$ is a point of discontinuity). The residual $U(1)$
symmetry are rotations about the isospin $Z$ axis which in the above
picture is obtained by rotation around the horizontal axis. In this
example the ``molecular'' field on the $k^{th}$ subsystem is
$\varepsilon (k)\left\langle \sigma _{k}^{z}\right\rangle +\sum
v(k,k^{\prime })\left\langle \sigma _{k^{\prime }}^{x}\right\rangle
$ if the $\left\langle b_{k}\right\rangle $'s are chosen real. The
zero mass is then the aforementioned rotation which for this ground
state are excitations which are linear combinations of $\sigma
_{k}^{y}$. There are also fermionic excitations which are massive.
Their mass corresponds to the energy necessary to break up a Cooper
pair. It is given by $\left[ \left( \varepsilon _{k}-\varepsilon
_{F}\right) ^{2}+{\it H} _{k}^{2}\right] ^{1/2}$ where ${\it
H}_{k}$ is the "transverse" molecular field on
$\vec{\sigma }_{k}$ \ given by $v(k,k^{\prime })\left[
\left\langle b_{k^{\prime }}\right\rangle +\left\langle b_{k^{\prime
}}^{+}\right\rangle \right] $.

In the above model the interaction $v(k,k^{\prime })$ is a small
attractive force that issues from exchange of phonons (lattice
waves) among the electrons. In addition there is a much stronger
force due to Coulomb interaction. Whereas Bardeen Cooper
Shrieffer \cite{bcs} worked only with the former,  Anderson
\cite{anderson} and Nambu \cite{nambu1} analysed the effects of the
latter. The fermionic mass is essentially unaffected, but the collective
mode is completely modified so as to become the massy plasmon. It is a
"longitudinal" photon. There are also massy ``transverse'' photons. These
give rise  to the Meissner effect and the flux tubes of type II
superconductors. The transverse and longitudinal masses are unequal
since their origins differ dynamically. The plasmon uses the total
electron density whereas the transverse photons refers to the
condensate (i.e. the
$\left\langle b_{k}\right\rangle $).

These effects were the precursor of the 
BEH mechanism which is studied in ``Modern Times''.  Then because there is
 longitudinal and transverse isotropy in the quantum relativistic quantum
field vacuum, there is  only one mass. 

\bigskip
{\bf \large VII. Spontanously Broken Chiral Symmetry(SB${\bf \chi }${\bf
S)}}
\medskip

One of the first exercises for students in field theory is the
calculation of the electron's self mass $\Delta m$, in QED with the
result to $O(e^{2})$
\begin{equation}\Delta m\sim e^{2}m_{0}\ln (\Lambda
/m_{0})\, ,
\end{equation}
where $m_{0}$ is the bare mass, $\Lambda $ the cut-off. The
important point is that $\Delta m=0$, if $m_{0}=0$. It is this
circumstance that reduces the divergence of $\Delta m$ from the
na\"{\i}ve expectation that is linear in $\Lambda $ to logarithmic.
One says that the mass is ``protected'' by chiral symmetry. Chiral
symmetry for a single fermion field is invariance of the action
under
\begin{equation}
\Psi \rightarrow e^{i\alpha \gamma _{5}}\Psi\, .
\end{equation}
Whereas under normal (global) gauge transformations the $L$ and $R$
components transform the same way (where $L,R=\left[ \left( 1\pm
\gamma _{5}\right) /2
\right] \Psi $) they transform with opposite signs under the
chiral gauge transformation. 

Since $\overline{\Psi }=\Psi ^{+}\gamma _{0}$ one has under Eq.(64)
$\overline{
\Psi }\rightarrow \overline{\Psi }e^{i\alpha \gamma _{5}}$. In
consequence
\begin{eqnarray}\overline{\Psi }\Psi &\rightarrow& \left( \cos
2\alpha \right) 
\overline{\Psi }\Psi +\left( \sin 2\alpha \right) \overline{\Psi
}i\gamma _{5}\Psi \nonumber\\
\overline{\Psi }i\gamma _{5}\Psi &\rightarrow& (-\sin 2\alpha
)\overline{\Psi  }i\gamma _{5}\Psi +(\cos 2\alpha )\overline{\Psi
}\Psi\, . \end{eqnarray}
 Here $\gamma _{5}$ is hermitian with $(\gamma
_{5})^{2}=1$ and $\left\{ \gamma _{5},\gamma _{\mu }\right\} =0$.
Thus under Eq.(65) the couple $\left( \overline{\Psi }\Psi
,\overline{\Psi } (i\gamma _{5})\Psi \right) $ transforms as a
vector under chiral transformations i.e. it rotates in the ``chiral
gauge plane'' with  angle ($ 2\alpha $).

Whereas the electromagnetic interaction, as well the kinetic term
in the action are chiral invariants (since $\left\{ \gamma
_{5},\gamma _{\mu }\right\} =0)$ thereby securing the invariance of
$\overline{\Psi }\gamma _{\mu }\Psi $, the mass term ($m_{0}\overline{\Psi
}\Psi $) is not, due to Eq.(65). One consequence is that every term in
pertubation theory gives $m=0$ if
$m_{0}=0$. This is easily checked by making the count of the number
of $
\gamma $  matrices appearing in vertices and fermion propagators. It
is odd and the trace of such a term vanishes. A mass term appearing
in the self energy is calculated by taking the trace. We shortly
give a more synthetic demonstration of this fact from the chiral
Ward identity.

Inspired by the BCS theory of superconductivity, wherein a mass gap
was derived non perturbatively (through Cooper bound state
formation), Nambu~\cite{nambu2} showed that the same could arise in
quantum field theory, the price being the existence of a dynamically
generated pseudoscalar meson, which he then identified with the pion.
During this same period, Gell-Mann and Levy \cite{gm} proposed a chiral
invariant action which contained scalar and pseudoscalar fields
coupled to the fermion (Yukawa coupling). SB$\chi $S was first
generated through the bosonic action (effective potential method)
wherein the scalar picked up an expectation value and the
pseudoscalar had zero mass in consequence of SBS kinematics.
At low momentum scales the two methods give equivalent physical
results, whereas at large momenta the composite character of the
effective boson fields in Nambu's methods could give a considerable
modification of the dynamics, so as to augment the width of the
massy scalar (i.e. the scalar which corresponds to the stretching
mode or longitudinal susceptibility in the magnetic case).

   It is the Gell-Mann L\'{e}vy phenomenological approach which until
the present time has prevailed in standard model research in the
implementation of the BEH mechanism. Research beyond the standard
model is so tenuous that all avenues must be considered open. One
also must bear in mind that the original dynamical mechanism of
SB$\chi $S of Nambu Jona-Lasinio is now supplanted by the QCD
confinement mechanism. In this case the zitterbewegung of quarks at
the end of electric flux tubes (the model for mesons) provides for
the ``constituent'' quark mass.  The chiral symmetry of QCD then
implies the existence of pions. These have zero mass if the
``current'' quark mass is zero and have a $(mass)^{2}$ proportional to
the latter when it is not zero. An exception is the ninth pseudoscalar of
the eightfold way which has mass due to an anomaly. The origin of quark
and lepton masses in terms of some ultimate chiral, super or GUTS
symmetry remains elusive, the Yukawa coupling in standard model
reseach being most likely the phenomenological expression of a deeper
theory.

In this review we shall adhere to Nambu's original approach since
it carries a pedagogical message of both power and elegance. We
first review the simple non perturbative approach of Nambu
Jona-Lasinio \cite{nambu3}, not that it need be of direct applicability, but
rather because it sets the scene for more general considerations.

Consider a chiral invariant four point interaction (such as $\lambda
\left[ 
\overline{\Psi }\gamma _{\mu }\Psi \right] ^{2}$ ) or its Fierz
equivalent $
\lambda \left[ (\overline{\Psi }\Psi )^{2}-(\overline{\Psi }\gamma
_{5}\Psi )
\right] ^{2}$. In lowest order the fermion self energy, $\Sigma $,
is given by the graph (Fig.7)

\hskip 3.5cm \epsfbox{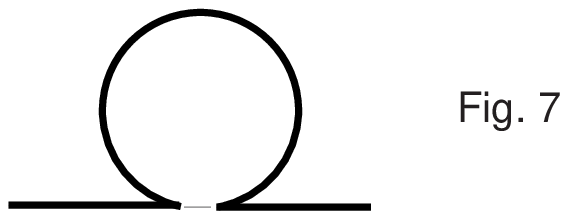}

where one may imagine some non locality over a distance $\Lambda
^{-1}$ at the vertex (say due to the exchange of a very heavy
meson) supplies a U-V cut-off. Then
 \begin{equation}\Sigma(p)\sim g\int^{\Lambda }\Gamma
\frac{d^{4}k}{ \gamma^\mu (k+p)_\mu}\Gamma \, ,
\end{equation}
where irrelevant factors of $O(1)$ are dropped and $\Gamma $ are the
relevant $\gamma $ matrices (e.g. take $\gamma _{\mu }$ for definiteness).
Then  $ tr\Sigma =0$ and $\Delta m=0$.

Let us now make this self consistent by iterating in the fermion
propagator so that
\begin{equation}\Sigma (p)\sim g\int^{\Lambda} d^{4}k\,  \Gamma
\frac{1}{\gamma^\mu (k+p)_\mu-\Sigma (k+p)}\Gamma\, .
\end{equation}

This corresponds to an infinite sum of graphs often called rainbow
graphs i.e a rainbow is built on every fermion line ad infinitum.
(It is amusing that one can build the Weiss molecular field of 
ferromagnetism using exactly the same set of graphs in a field
theory which is equivalent to the original spin problem).

Let $\Sigma =A(p)\gamma^\mu p_\mu +M(p)$  and take the trace to give
\begin{equation}M(p)\sim g\int^{\Lambda
}d^{4}k\, \frac{M(p+k)}{(p+k)^{2}-M^{2}(p+k)}\, ,
\end{equation}
where we have not included the effect due to the form factor $A$. The
ensuing integral equation is difficult to solve but the ideas are
brought out setting $M(p)=M$= constant so as to give an eigenvalue
equation for $M$. Taking into account factors for $i$, one gets a
solution by making a Wick rotation provided the force is attractive
($g< 0$). This is the equivalent to the gap equation in
superconductors.

Of particular interest in particle theory is the accompanying
pseudoscalar. Note the SB$\chi $S; one could have taken $M$ as a
linear combination of $ M_{1}+iM_{2}\gamma_5$  with
$M^{2}=M_{1}^{2}+M_{2}^{2}$. Choosing $M_2=0$ is a choice of direction
in the chiral gauge plane, along the axis $\overline{\Psi }\Psi 
$. Then $\overline{\Psi }\gamma _{5}\Psi $ should propagate with
zero mass. It does as seen from its propagator  ~~~ ($\sim 1/1-
g\Pi$) (Fig 8)
\vskip.5cm

 \hskip 1cm  \epsfbox{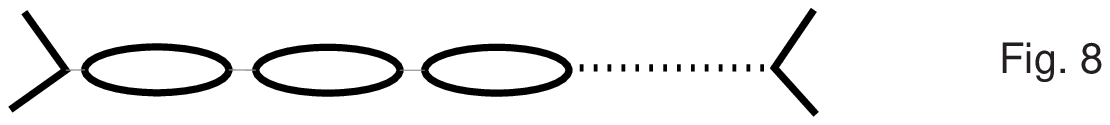}

\begin{equation}
\Pi(p)=tr\int d^4k\,  \Gamma \frac{1}{\gamma^\mu(p+k)_\mu-M}\Gamma
\frac{1}{\gamma^\mu k_\mu-M}\, .
\end{equation}

At $q=0$ one has
\begin{equation}
\Pi (0)\simeq \int
\frac{d^{4}k}{k^{2}-M^{2}}\, ,
\end{equation}
and from the eigenvalue condition Eq.(68) one checks $1-g
\Pi(0)=0$. The diligent reader may check
all the kinematic factors of $O(1)$. Thus the propagator of
$\overline{\Psi }\gamma _{5}\Psi $ at $q=0$ has a pole and one may
check (for example by dispersion relations) that it corresponds to
a pole at $q^{2}=0$ in the more general case when $q_{\mu }\neq 0$. ( Here
$q^{2}\equiv q_{0}^{2}-\underline{q}^{2}$\ )

This result is general, powerful and independent of approximations
that have been made. That is the true powerful accomplishement of
Nambu which we now present.

The chiral Ward identity, established in the  same way as the
usual vector Ward identity through use of the symmetry of the
action under chiral transformations, reads
\begin{equation}\lim_{q_{\mu }\rightarrow 0}q_{\mu }\Gamma
_{\mu 5}=\gamma _{5}\Sigma(p+q)+\Sigma (p)\gamma _{5}\, .
\end{equation}

$\Gamma _{\mu 5}$ is the vertex function formed from the chiral
sources of momentum $q_{\mu }$\ which scatters a fermion from
$p_{\mu }$ to $p_{\mu }+q_{\mu }$. As $q_{\mu }\rightarrow 0$,
one sees that the form factor $A(p)$ drops out of Eq.(71) (since
$\left\{ \gamma _{5},\gamma _{\mu }\right\} =0
$).
Whence
\begin{equation}
\lim_{q_{\mu }\rightarrow 0}q_{\mu }\Gamma _{\mu
5}=2M(p)\gamma _{5}\quad ,\quad \Gamma _{\mu 5}\rightarrow \frac{2M(p)\gamma
_{5}q_{\mu }}{q^{2}}\, .
\end{equation}
This pole at $q^2=0$ is the signal of a pseudoscalar meson which
couples to the fermion field through the mass of the fermion.

Nambu recognized that in this way he had discovered the key to the
success of the celebrated Goldberger-Treiman relation, one of the
gems of particle physics in the 1950's-1960's, to which we now turn
so closing out this review of the Paleolithic Age.

The original derivation by Goldberger and Treiman was based on a
dispersion relation argument, involving two assumptions: an
unsubtracted dispersion relation for one of the form factors
occurring in the matrix element for $
\beta $ decay (see below) and pion dominance of the same. The
quantitative success was remarkable, but there was little
understanding of how to justify the assumptions. This was supplied
by Nambu as follows. Whereas the Ward identity involves fields, one
can also work directly with matrix elements of currents among
physical states. In particular the matrix element of the axial
current between nucleons is observed as the Gamow-Teller transition
in 
$\beta $ decay. Its most general form can be shown to be
 \begin{equation}\left\langle N\left\vert j_{\mu
5}\right\vert N^{\prime }\right\rangle
=F_{A}(q^{2})\overline{u}_{N}(p+q)\gamma _{\mu }\gamma
_{5}u_{N^{\prime }}(p)+F_{p}(q^{2})\overline{u}_{N}(p+q)q_{\mu
}\gamma _{5}u_{N}(p)\, ,
\end{equation}
where $q_{\mu }$ is the 4-momentum tranfer carried by $j_{\mu 5}$.

Let us suppose that $\partial _{\mu }j_{\mu 5}=0$ (i.e. chiral
invariance). Then taking the divergence of Eq.(73) gives
 \begin{equation}0=\left[ (m_{N}+m_{N^{\prime
}})F_{A}(q^{2})+q^{2}F_{p}(q^{2})
\right] \left[ \overline{u}_{N}(p+q)\gamma _{5}u_{N}(p)\right]\, ,
\end{equation}
where we have used $\left\{ \gamma _{5},\gamma _{\mu }\right\} =0$
and the Dirac equation $(\gamma^\mu p_\mu -M)\Psi =0$. As
$q^{2}\rightarrow 0$, one finds
\begin{equation}F_{p}(q^{2})\to_{q^{2}\rightarrow
0}(m_{N}+m_{N^{\prime }})\frac{1}{q^{2}}F_{A}(0)\, ,
\end{equation}
i.e. $F(p)$ has a pole which like $\Gamma _{\mu 5}$ (Eq.(72)) has
residue proportional to the fermion mass, here the nucleon. Eqs.
(74) and (75) have the interpretation given by the graphical structure
for $F_p$ (Fig.9)

\bigskip 

\hskip 2.5cm\epsfbox{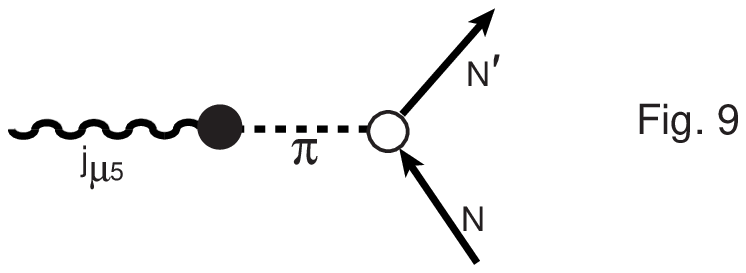}
\bigskip

The dark circle in the drawing is $\left\langle 0\left\vert j_{\mu 5}
\right\vert \pi \right\rangle = -i f_\pi q_\mu$. The dotted line
is the pion propagator ($=1/q^{2}$) and the light circle is $g_{\pi
NN^{\prime }}
\overline{u}_{N^{\prime }}\gamma _{5}u_{N}$, where $g_{\pi
NN^{\prime }}$ is the pion nucleon coupling constant. The residue
condition implied by Eq.(75) is
 \begin{equation}f_{\pi }g_{\pi NN^{\prime
}}=(m_{N}+m_{N^{\prime }})F_{A}\, .
\end{equation}

All quantities are measured; $f_{\pi }$ is the pion decay constant
into leptons, $F_{A}$ the Gamow-Teller $\beta $\ decay constant,
$g_{\pi NN^{\prime }}$ is found from $\pi N$ scattering. Agreement
is found to $O(2\%) $. The deviation is attributed to the fact that
$\partial _{\mu }j_{\mu 5}$ does not quite vanish and this is
reflected by the fact that $m_{\pi }^{2}\neq 0$. Rather 
 \begin{equation}\left\langle 0\left\vert \partial _{\mu
}j_{\mu 5}\right\vert \pi
\right\rangle =q^{2}f_{\pi }=m_{\pi }^{2}f_{\pi }\, ,
\end{equation}
since the pion is ``on shell''. Nambu attributed $m_{\pi }^{2}\neq 0$, (but
small on the scale of hadron physics) to a small external breaking of chiral
invariance induced by a bare mass, $m_{0}$. He then showed that $m_{\pi
}^{2}\sim m_{0}M$ where $M$= hadron scale ($m_{0}\sim 10Mev,M\sim
1Gev,m_{\pi }\sim 130Mev$). Note the analogy to Eq.(55).

Now return to Eqs.(73), (74) and (75). The l.h.s. of Eq.(74) now contains
the non vanishing value $\left\langle N\left\vert \partial _{\mu
}j_{\mu 5}\right\vert N^{\prime }\right\rangle $. Nevertheless the
residue relation Eq.(76) should not change significantly. The
$(momentum~transfer)^{2}$ in $g_{\pi NN^{\prime }}$\ and in $F_{A}$
are now shifted by $O(m_{\pi }^{2})$. Therefore Eq.(76) ought to
hold good at the 1\% level. The corrections will be encoded in the
contribution of high mass states not included in the pion dominance
estimate of $F_{p}$. Assuming this true, one replaces Eq.(75) by
\begin{equation}F_{p}(q)\simeq (m_{N}+m_{N^{\prime
}})\frac{1}{q^{2}-m_{\pi }^{2}}
 F_{A}(0)\, ,
\end{equation}
 then yields Eq (76). This is the famous principle of PCAC
wherein pion matrix elements are related to matrix elements of the
axial current. See reference \cite{weinberg} for the phenomenology
development of soft pion physics. When united with the Gell-Mann current
algebra it becomes a very powerful tool which interrelates all kinds of
hadronic phenomena, thus becoming one of the dominant elements of
particle physics throughout the 1960's and early 70's. The success of the
whole development bit by bit led to the QCD quark model and
confinement which are now considered the theoretical bases of hadron
physics as well as hadron-lepton interactions.

In the preceedings paragraphs, we have seen the important role that SBS
has played in particle physics when the symmetry that has been broken
is global (the chiral group). ``Modern Times'' is devoted to the other
important facet of this development, to wit: the BEH mechanism wherein
one adds to the previous consideration the complication of gauge
symmetry. This chapter lead to the electroweak unification.

 

\end{document}